\documentclass[journal]{IEEEtran}
\ifdefined\pdfminorversion\pdfminorversion=4\fi
\usepackage{cite,amsmath,amssymb,mathtools,bm,graphicx,booktabs,microtype,balance,placeins}
\usepackage[hypertexnames=false]{hyperref}
\hypersetup{hidelinks,
 pdftitle={Prescribed-Performance Tracking Under Limited and Degraded Actuation},
 pdfauthor={Yutong Dong, Zongyi Guo, and Thomas Berger}}
\newtheorem{assumption}{Assumption}
\newtheorem{proposition}{Proposition}
\newtheorem{theorem}{Theorem}
\newtheorem{corollary}{Corollary}
\newcommand{\T}{^{\mathrm{T}}}

\newcommand{\R}{\mathbb R}
\newcommand{\bp}{\bm p}\newcommand{\bv}{\bm v}\newcommand{\bu}{\bm u}
\newcommand{\bd}{\bm d}\newcommand{\bff}{\bm f}\newcommand{\bB}{\bm B}
\newcommand{\bI}{\bm I}\newcommand{\bx}{\bm x}\newcommand{\bfo}{\bm f_0}
\newcommand{\bep}{\bm e_p}\newcommand{\bev}{\bm e_v}\newcommand{\bpd}{\bm p_d}
\newcommand{\balpha}{\bm\alpha}\newcommand{\bzp}{\bm z_p}\newcommand{\bzv}{\bm z_v}
\newcommand{\D}{\mathcal D}\newcommand{\sym}{\operatorname{sym}}\newcommand{\proj}{\operatorname{proj}}
\newcommand{\FigRef}[1]{\hyperref[#1]{\textbf{Fig.~\ref*{#1}}}}
\newcommand{\TabRef}[1]{\hyperref[#1]{\textbf{Table~\ref*{#1}}}}
\begin{document}
\title{Prescribed-Performance Tracking Under Limited and Degraded Actuation}
\author{Yutong Dong, Zongyi Guo, and Thomas Berger\thanks{This work was supported in part by the National Natural Science Foundation of China under Grant 52272404 and by the Natural Science Basic Research Program of Shaanxi under Grant 2026JC-YXQN-225. T. Berger acknowledges funding by the Deutsche Forschungsgemeinschaft (DFG, German Research Foundation), Project-ID 544702565. (Corresponding author: Zongyi Guo.)}\thanks{Y. Dong is with the Department of Industrial and Systems Engineering, The Hong Kong Polytechnic University, Hong Kong, China (e-mail: yutong.dong@polyu.edu.hk).}\thanks{Z. Guo is with the School of Astronautics, Northwestern Polytechnical University, Xi'an, China (e-mail: guozongyi@nwpu.edu.cn).}\thanks{T. Berger is with the Institut f{\"u}r Mathematik, Martin-Luther-Universit{\"a}t Halle-Wittenberg, Theodor-Lieser-Stra{\ss}e~5, 06120 Halle, Germany (e-mail: thomas.berger@mathematik.uni-halle.de).}}
\maketitle
\begin{abstract}
Actuator degradation can prevent prescribed tracking under a fixed input limit. For multivariable systems, we derive conditions for projected radial feedback to retain position and auxiliary-velocity bounds during partial loss and restore the auxiliary reserve. An onset estimate allows bounded gain perturbations. During complete loss, a position estimate gives a duration for maintaining the bound. A scalar braking calculation compares the minimum plant input with a sufficient feedback threshold. Spacecraft simulations illustrate these results.
\end{abstract}
\begin{IEEEkeywords}
Actuator degradation, input constraints, nonlinear systems, prescribed performance, robust control.
\end{IEEEkeywords}

\section{Introduction}

\IEEEPARstart{P}{rescribed} performance control (PPC) confines tracking errors within bounds selected in advance. Their decay rates and final values specify the desired transient and steady-state response \cite{Bechlioulis2008,Bechlioulis2009}. Funnel control pursues a related objective \cite{Ilchmann2002}. Both methods have been applied to uncertain multivariable systems and systems with actuator faults \cite{Dimanidis2020,Zhao2023}.

Input limits have motivated reference modification \cite{Fotiadis2024}, performance bounds that change with operating conditions \cite{Yong2020,Trakas2024}, and switching controllers \cite{Bikas2024}. Berger \cite{Berger2024} developed a funnel controller whose boundary widens during saturation and returns to its prescribed shape afterward. Bikas et al.\ \cite{Bikas2026} considered prescribed tracking together with state constraints.

If actuator effectiveness has a uniform positive lower bound and the command is unbounded, $\eta(t)\bB(t,\bx)$ can be treated as the input gain. Existing PPC and funnel-control methods then apply under their respective assumptions. With a finite input limit, decreasing effectiveness leaves less control available to maintain a fixed tracking bound. During complete loss, the commanded input has no effect on the plant acceleration. The state at loss onset, drift, and disturbance determine how long the position bound can be maintained. With residual actuation, the auxiliary velocity bound imposes a further input requirement.

\begin{figure}[!t]\centering
\includegraphics[width=.95\columnwidth]{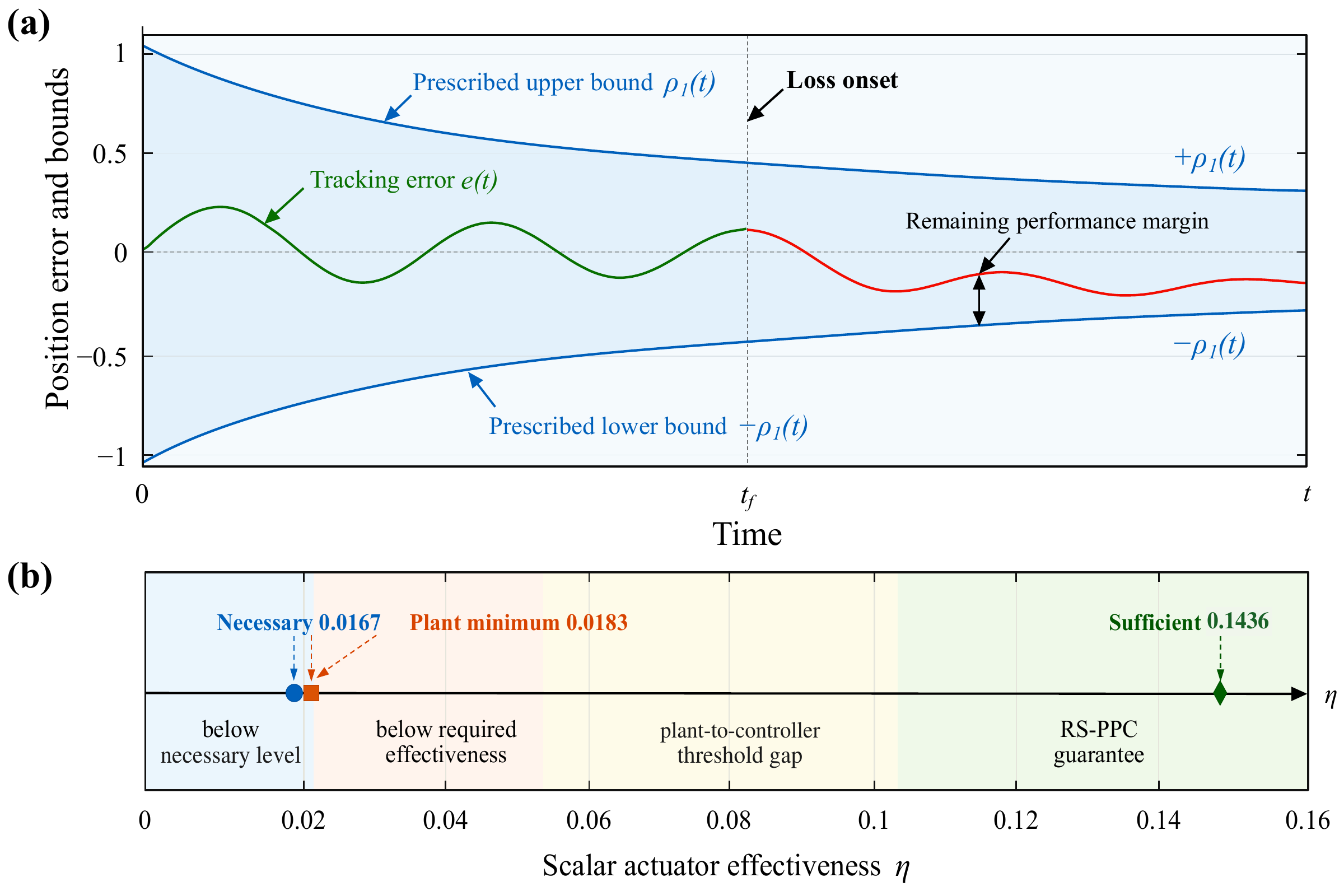}
\caption{(a) Tracking margin during actuator loss. (b) Effectiveness levels for the nominal scalar example.}\label{fig:framework}
\end{figure}

Set invariance provides a way to determine the control required to maintain a physical constraint \cite{Blanchini1999,Aubin1991,Miani2015}. We compare that requirement with conditions for the projected radial feedback. The two contributions are as follows.
\begin{enumerate}
\item \emph{Prescribed bounds under degraded actuation.} We derive rates for the position and auxiliary-velocity reserves. Their integrals give conditions for retaining both error bounds during partial loss and for restoring the auxiliary reserve. A bound starting from the state at loss onset also allows bounded perturbations of the input gain.
\item \emph{Input required by the plant and the feedback.} During complete loss, we bound how long the physical position constraint remains satisfied. For a scalar system, a braking calculation gives the minimum input required by the plant. We compare it with a sufficient level for a specified family of radial feedbacks.
\end{enumerate}
\FigRef{fig:framework} illustrates the remaining position margin and the scalar input requirements. Spacecraft simulations compare the bounds with the observed boundary contacts and tracking response.

\section{System and Controller}\label{sec:model}
\subsection{Model and Bounds}
Consider the multivariable system
\begin{equation}
 \dot\bp=\bv,\qquad
 \dot\bv=\bff(t,\bx)+\eta(t)\bB(t,\bx)\bu+\bd(t),
 \label{eq:plant}
\end{equation}
where $\bx=\operatorname{col}(\bp,\bv)\in\R^{2m}$, $\bp,\bv\in\R^m$ are position and velocity, and $\bu\in\R^m$ is the commanded input. The disturbance is $\bd\in\R^m$. We impose
\begin{equation}
 \|\bu\|\le U,\qquad 0\le\eta_L(t)\le\eta(t)\le\bar\eta,
 \label{eq:limits}
\end{equation}
with known $U>0$, $\eta_L$, and $\bar\eta$. The common effectiveness $\eta$ scales all input channels, giving the effective input $\eta\bu$ and acceleration $\eta\bB\bu$. Complete loss corresponds to $\eta=0$.

\begin{assumption}[Unknown dynamics]\label{ass:model}
On an open domain $\mathcal X$, the functions $\bff$, $\bB$, and a known nominal drift $\bfo$ are measurable in time and locally Lipschitz in state. Their values and state Lipschitz moduli are locally bounded on finite time intervals. On the state neighborhoods used below, known bounds satisfy
\begin{equation}
\begin{gathered}
 \|\bff-\bfo\|\le\bar\Delta_f,\qquad \|\bd\|\le\bar d,\\
 \sym(\bB)\succeq b_-\bI_m,\qquad\|\bB\|\le b_+,
\end{gathered}
 \label{eq:model_bounds}
\end{equation}
where $0<b_-\le b_+$, $\sym(\bB)=(\bB+\bB\T)/2$, and $\bI_m$ is the identity matrix. The nonnegative envelopes $\bar\Delta_f(t,\bx)$ and $\bar d(t)$ have the same local boundedness and state regularity. The signals $\bd$ and $\eta$ are measurable. Position and velocity are measured.
\end{assumption}
These bounds supply the input contribution and the uncertain acceleration terms used in the margin estimates below.

\begin{assumption}[Reference and prescribed bounds]\label{ass:reference}
The reference belongs to $W^{2,\infty}_{\rm loc}$, and the positive radii $\rho_1,\rho_2$ belong to $W^{1,\infty}_{\rm loc}$. Their values have positive lower bounds on each finite interval under consideration. The reference and its first two derivatives, and the radii and their first derivatives, are available almost everywhere.
\end{assumption}
Here $W^{k,\infty}_{\rm loc}$ denotes locally essentially bounded weak derivatives through order $k$. We use continuous representatives of $\bpd$, $\dot\bpd$, and the radii, with derivatives understood almost everywhere.

All norms are Euclidean or induced matrix norms. Write $[s]_+=\max\{s,0\}$. The position requirement is
\begin{equation}
 \|\bep(t)\|<\rho_1(t),\qquad \bep=\bp-\bpd,
 \label{eq:objective}
\end{equation}
with a radius prescribed independently of the actuation history. Define
\begin{equation}
 \delta(t,\bx)=\bar\Delta_f(t,\bx)+\bar d(t),
 \label{eq:delta}
\end{equation}
so that $\|\bff-\bfo+\bd\|\le\delta$. Setting $\bfo=\bm0$ and $\bar\Delta_f=\bar f$ gives an implementation based on a known drift bound. 

\subsection{Projected Radial Feedback}\label{sec:controller}
The controller uses
\begin{equation}
\begin{aligned}
 D_p&=\max\{\rho_1^2-\|\bep\|^2+\epsilon_p,\epsilon_p/4\},
 &\bzp&=\bep/D_p,\\
 \balpha&=\dot\bpd-k_p\bzp,& \bev&=\bv-\balpha,\\
 D_v&=\max\{\rho_2^2-\|\bev\|^2+\epsilon_v,\epsilon_v/4\},
 &\bzv&=\bev/D_v,
\end{aligned}\label{eq:coordinates}
\end{equation}
where $k_p>0$, $\epsilon_p,\epsilon_v>0$, and $\balpha$ is a virtual velocity. Let
\begin{equation}
 r_p=1-\|\bep\|^2/\rho_1^2,
 \qquad r_v=1-\|\bev\|^2/\rho_2^2.
 \label{eq:reserves}
\end{equation}

The conditions $r_p>0$ and $r_v>0$ express the position and auxiliary-velocity error bounds, respectively.

On $r_p\ge0$, differentiation gives
\begin{equation}
\begin{aligned}
 \dot\bzp&=\frac{\bv-\dot\bpd}{D_p}
 -\frac{2\bep[\rho_1\dot\rho_1-\bep\T(\bv-\dot\bpd)]}{D_p^2},\\
 \dot\balpha&=\ddot\bpd-k_p\dot\bzp.
\end{aligned}\label{eq:alphadot}
\end{equation}

The derivative $\dot\balpha$ uses the measured state and the first two reference derivatives. To account for its contribution to the velocity error, choose $\rho_{2,w}=\beta\rho_2$, with $0<\beta<1$, and set
\begin{equation}
\begin{aligned}
 \gamma_v&=\frac{\rho_{2,w}}{\rho_2^2-\rho_{2,w}^2+\epsilon_v},
 &s_v&=\frac{\gamma_v}{\sqrt{\gamma_v^2+\nu_v^2}},\\
 \Psi_v&=\|\bfo-\ddot\bpd\|+\delta+\|\ddot\bpd-\dot\balpha\|,\\
 H_v&=\Psi_v+|\dot\rho_{2,w}|+\delta_v,
\end{aligned}\label{eq:compensator}
\end{equation}
where $\nu_v,\delta_v>0$. The input is
\begin{equation}
 \bu=\proj_{\mathbb B_U}\left[-m_*\left(k_v\bzv+
 \frac{H_v}{s_v}\frac{\bzv}{\sqrt{\|\bzv\|^2+\nu_v^2}}\right)\right],
 \label{eq:controller}
\end{equation}
where $k_v,m_*>0$, $\mathbb B_U=\{\bu:\|\bu\|\le U\}$, and $\proj$ denotes Euclidean projection. For $q_v=\|\bev\|>0$, this command is $\bu=-a_v\bev/q_v$ with
\begin{equation}
 a_v=\min\left\{U,m_*\left(k_vh_v+
 \frac{H_vh_v}{s_v\sqrt{h_v^2+\nu_v^2}}\right)\right\},
 \quad h_v=\|\bzv\|.
 \label{eq:amplitude}
\end{equation}
At $q_v=0$, both $a_v$ and the command are zero. The candidate region $\D(t)=\{(\bp,\bv):r_p,r_v\ge0\}$ satisfies
\begin{equation}
 \|\bp\|\le\|\bpd\|+\rho_1,
 \quad\|\bv\|\le\|\dot\bpd\|+k_p\rho_1/\epsilon_p+\rho_2.
 \label{eq:state_caps}
\end{equation}

On a neighborhood of the candidate region, the regularized coordinates are locally Lipschitz and their denominators are bounded away from zero. Together with the projection and Assumptions~\ref{ass:model}--\ref{ass:reference}, this yields a unique local Carath\'eodory solution. 

\section{Main Results}\label{sec:results}
\subsection{Position Bound During Complete Loss}\label{sec:loss}
For $\eta=0$, the physical error satisfies
\begin{equation}
 \ddot{\bep}=\bff+\bd-\ddot{\bpd}.
 \label{eq:outage}
\end{equation}
Let $t_f$ denote the beginning of complete loss. Set $\bm e_f=\bep(t_f)$ and $\bm\nu_f=\bv(t_f)-\dot\bpd(t_f)$. To bound their subsequent motion, choose an integrable nonnegative function $M$ such that
\begin{equation}
 \|\bfo(t,\bx)-\ddot\bpd(t)\|+\delta(t,\bx)\le M(t)
 \label{eq:acc_bound}
\end{equation}
throughout a compact neighborhood $\mathcal K\subset\mathcal X$. Define
\begin{equation}
 J(\tau)=\int_0^\tau(\tau-s)M(t_f+s)\,ds,
 \quad K(\tau)=\int_0^\tau M(t_f+s)\,ds.
 \label{eq:outage_integrals}
\end{equation}
For $\mathbb B(\bm c,r)=\{\bm z:\|\bm z-\bm c\|\le r\}$, the associated enclosures are
\begin{equation}
\begin{aligned}
 \bp(t_f+\tau)&\in\bpd(t_f+\tau)
                  +\mathbb B(\bm e_f+\tau\bm\nu_f,J(\tau)),\\
 \bv(t_f+\tau)&\in\dot\bpd(t_f+\tau)
                  +\mathbb B(\bm\nu_f,K(\tau)).
\end{aligned}\label{eq:outage_enclosure}
\end{equation}
We require the position and velocity enclosures jointly to remain in the interior of $\mathcal K$ over the interval considered. 

\begin{theorem}[Position bound during complete loss]\label{thm:outage}
Suppose the stated drift and reference regularity conditions hold on $[t_f,t_f+T]$, with $\eta=0$. If the enclosures satisfy the stated inclusion and
\begin{equation}
 \|\bm e_f+\tau\bm\nu_f\|+J(\tau)<\rho_1(t_f+\tau)
 \quad (0\le\tau\le T),
 \label{eq:outage_test}
\end{equation}
 then the solution continues through $t_f+T$ and satisfies the physical position bound throughout that interval.
\end{theorem}
\begin{IEEEproof}
We integrate the error acceleration up to the first possible exit from $\mathcal K$:
\begin{equation}
 \bep(t_f+\tau)=\bm e_f+\tau\bm\nu_f
 +\int_0^\tau(\tau-s)\ddot\bep(t_f+s)\,ds.
 \label{eq:outage_identity}
\end{equation}
The acceleration bound yields both enclosures in \eqref{eq:outage_enclosure}. At a first exit from $\mathcal K$, the state would reach its boundary, contrary to the assumed strict inclusion of the enclosures. Thus the state remains in $\mathcal K$ throughout $[t_f,t_f+T]$.

Condition~\eqref{eq:outage_test} keeps the position error strictly inside $\rho_1$. The compact enclosure and local regularity allow the solution to continue through $t_f+T$.
\end{IEEEproof}
A conservative scalar test uses $E_f=\|\bm e_f\|$ and $V_f=\|\bm\nu_f\|$. If $M\le M_*$ and $-\dot\rho_1\le L_\rho$, with $M_*,L_\rho\ge0$, it is enough that
\begin{equation}
 (V_f+L_\rho)T+\tfrac12M_*T^2<m_f,
 \quad m_f=\rho_1(t_f)-E_f>0.
 \label{eq:loss_quadratic}
\end{equation}

The quantity $m_f$ is the distance remaining to the position boundary at loss onset. For the acceleration inclusion $\ddot\bep\in\mathbb B(\bm0,M(t))$, the ball in \eqref{eq:outage_enclosure} contains every position reachable at a fixed time. Its largest error magnitude gives the test in \eqref{eq:outage_test}.

\subsection{Margin Rates and Preservation}
\begin{proposition}[Rates from the available model data]\label{prop:rates}
On $\D(t)$, the feedback satisfies
\begin{equation}
\begin{aligned}
 \dot r_p=L_p={}&\frac{2k_p\|\bep\|^2}{\rho_1^2D_p}
 -\frac{2\bep\T\bev}{\rho_1^2}
 +\frac{2\|\bep\|^2\dot\rho_1}{\rho_1^3},\\
 \dot r_v\ge L_v^{\rm r}={}&\frac{2}{\rho_2^2}
 [\eta_L b_-q_va_v-\bev\T(\bfo-\dot\balpha)-q_v\delta]
 +\frac{2q_v^2\dot\rho_2}{\rho_2^3},\\
 L_v^{\rm r}\ge L_v^{\rm n}={}&\frac{2q_v}{\rho_2^2}
 (\eta_Lb_-a_v-\Psi_v)+\frac{2q_v^2\dot\rho_2}{\rho_2^3}.
\end{aligned}\label{eq:rates}
\end{equation}
The superscripts ${\rm r}$ and ${\rm n}$ distinguish the directional estimate from its norm bound. Both rates vanish at $q_v=0$.
\end{proposition}
\begin{IEEEproof}
Differentiating $r_p$ along $\dot\bep=\bev-k_p\bzp$ gives the position rate. For the velocity reserve, write $\bm\omega=\bff-\bfo+\bd$, with $\|\bm\omega\|\le\delta$. We then obtain
\begin{equation}
 \dot r_v=-\frac{2\bev\T(\bfo-\dot\balpha+\bm\omega+\eta\bB\bu)}{\rho_2^2}
 +\frac{2q_v^2\dot\rho_2}{\rho_2^3}.
 \label{eq:rv_identity}
\end{equation}
The radial command opposes $\bev$. From $\sym(\bB)\succeq b_-\bI_m$, we obtain $-\bev\T\bB\bu\ge b_-q_va_v$. The residual contributes at least $-q_v\delta$. These bounds in \eqref{eq:rv_identity} yield $L_v^{\rm r}$. Finally, $\bev\T(\bfo-\dot\balpha)\le q_v(\Psi_v-\delta)$ gives $L_v^{\rm n}$.
\end{IEEEproof}
At a fixed state with $q_v>0$, we next compare the applied amplitude $a_v$ with the input limit $U$. Set $N_v=\bev\T(\bfo-\dot\balpha)+q_v\delta-q_v^2\dot\rho_2/\rho_2$. Let $\mathcal U$ contain all independent combinations allowed by the residual, gain, and effectiveness bounds in \eqref{eq:model_bounds} and \eqref{eq:limits}. The largest worst-case reserve rate is
\begin{equation}
\begin{aligned}
 L_v^{\max}&:=\max_{\|\bu\|\le U}\inf_{\mathcal U}\dot r_v
       =\frac{2}{\rho_2^2}(\eta_Lb_-q_vU-N_v),\\
 L_v^{\max}-L_v^{\rm r}&=\frac{2\eta_Lb_-q_v}{\rho_2^2}(U-a_v).
\end{aligned}\label{eq:authority_rate}
\end{equation}
Taking $\bB=b_-\bI_m$, $\eta=\eta_L$, and $\bm\omega=\delta\bev/q_v$ gives the worst-case upper bound for every $\|\bu\|\le U$. The radial choice $\bu=-U\bev/q_v$ attains that bound for the admitted uncertainty class. This establishes the first identity in \eqref{eq:authority_rate}. Replacing $U$ by the applied amplitude $a_v$ gives the second identity and identifies the remaining input capacity.

To determine the contribution of $H_v$, compare the two amplitudes at the same state. With the compensation term set to zero, $a_v^0=\min\{U,m_*k_vh_v\}$. Their difference is
\begin{equation}
\begin{aligned}
 \Delta a_v&=a_v-a_v^0
 =\min\left\{U-a_v^0,\frac{m_*H_vh_v}{s_v\sqrt{h_v^2+\nu_v^2}}\right\},\\
 L_v^{\rm r}-L_v^{\rm r,0}
 &=\frac{2\eta_Lb_-q_v}{\rho_2^2}\Delta a_v\ge0.
\end{aligned}\label{eq:compensation_gain}
\end{equation}
For $\eta_L>0$ and $q_v>0$, the compensation increases the lower rate whenever the basic command remains below $U$ and the added amplitude is positive. Once $a_v^0=U$, the input limit fixes the rate contribution.

Choose $r_p^*,r_v^*\in(0,1)$. The set $0\le r_j\le r_j^*$ is the layer used to assess contact with boundary $j$. On an interval $I$, let $\rho_{i,-}=\inf_I\rho_i$, $\rho_{i,+}=\sup_I\rho_i$, and $d_{\rho,i}=\operatorname*{ess\,sup}_I[-\dot\rho_i]_+$. With $\overline\Psi_v$ bounding $\Psi_v$ over the candidate region, retain the simple constants
\begin{equation}
\begin{aligned}
 h_{v,*}&=\frac{\rho_{2,-}\sqrt{1-r_v^*}}{\rho_{2,+}^2r_v^*+\epsilon_v},
 &a_{v,*}&=\min\{U,m_*k_vh_{v,*}\},\\
 G_p&=\frac{2k_p(1-r_p^*)}{\rho_{1,+}^2r_p^*+\epsilon_p},
 &A_p&=\frac{2(\rho_{2,+}+d_{\rho,1})}{\rho_{1,-}},\\
 G_v&=\frac{2b_-a_{v,*}\sqrt{1-r_v^*}}{\rho_{2,+}},
 &A_v&=\frac{2(\overline\Psi_v+d_{\rho,2})}{\rho_{2,-}}.
\end{aligned}\label{eq:uniform}
\end{equation}
On the corresponding layers, the constants yield
\begin{equation}
 \dot r_p\ge G_p-A_p,\qquad \dot r_v\ge\eta_LG_v-A_v.
 \label{eq:layer_rates}
\end{equation}
For the position layer, $\|\bep\|^2/\rho_1^2\ge1-r_p^*$ and $D_p\le\rho_{1,+}^2r_p^*+\epsilon_p$. On the velocity layer, $h_v\ge h_{v,*}$ and $\sqrt{1-r_v^*}/\rho_{2,+}\le q_v/\rho_2^2\le1/\rho_{2,-}$. Inserting these bounds into \eqref{eq:rates} gives the inward and adverse terms in \eqref{eq:layer_rates}.

\paragraph{Loss and reserve recovery}\label{subsec:recovery}
Let $\ell_p,\ell_v$ be integrable lower bounds for $L_p,L_v^{\rm r}$ on their respective layers. The velocity estimate uses the applied amplitude in \eqref{eq:amplitude}. To measure how much of either reserve may be lost, define
\begin{equation}
 R_j(t)=\min\{r_j(t),r_j^*\},\qquad
 Q_j(t,t_a)=\int_{t_a}^t[-\ell_j(s)]_+\,ds,
 \label{eq:Q}
\end{equation}
where $j=p,v$ and $R_j$ is the capped reserve.
\begin{theorem}[Preservation of both bounds]\label{thm:two}
Under Assumptions~\ref{ass:model}--\ref{ass:reference}, suppose $r_p(t_a),r_v(t_a)>0$ and the candidate sets over $[t_a,t_b]$ lie in a compact subset of $\mathcal X$. If $Q_j(t_b,t_a)<R_j(t_a)$ for $j=p,v$, the solution continues uniquely through $t_b$ and
\begin{equation}
 R_j(t)\ge R_j(t_a)-Q_j(t,t_a)>0,
 \quad t\in[t_a,t_b].
 \label{eq:two_bound}
\end{equation}
\end{theorem}
\begin{IEEEproof}
Consider the solution before either reserve first reaches zero. The functions $r_j$ and $R_j=\min\{r_j,r_j^*\}$ are absolutely continuous. Below the cap, the layer estimate gives $\dot R_j\ge\ell_j$. Above the cap, $R_j$ is constant, and its derivative is zero almost everywhere on the level set $r_j=r_j^*$. Thus $\dot R_j\ge-[-\ell_j]_+$ almost everywhere. Integrating from $t_a$ gives \eqref{eq:two_bound}. The assumed bound on $Q_j$ keeps its right-hand side positive, excluding a first zero before $t_b$. The state bounds in \eqref{eq:state_caps}, together with the compact-domain condition, permit continuation through $t_b$ \cite{Khalil2002}.
\end{IEEEproof}
Suppose $G_p\ge A_p$. For an integrable lower velocity rate $b(t)$, define
\begin{equation}
\begin{aligned}
 S(t)&=R_v(t_a)+\int_{t_a}^t b(s)\,ds,\\
 z(t)&=S(t)-\max\{0,\sup_{t_a\le s\le t}[S(s)-r_v^*]\}.
\end{aligned}\label{eq:reflection}
\end{equation}
The term subtracted from $S$ limits the recovered reserve to $r_v^*$.
\begin{corollary}[Loss duration and reserve recovery]\label{cor:recovery}
Under the initial-state, regularity, and domain conditions of Theorem~\ref{thm:two}, suppose $G_p\ge A_p$. If $z>0$ on the interval, then $R_v\ge z$ and both bounds are retained. For $0\le\eta_s\le\eta_L$ and $b=G_v\eta_s-A_v<0$, it is enough that
\begin{equation}
 T_{\rm loss}<\frac{R_v(t_f)}{A_v-G_v\eta_s}.
 \label{eq:loss_time}
\end{equation}
Starting with both reserves positive, a rate $b\ge\sigma_v>0$ restores the auxiliary reserve cap within
\begin{equation}
 T_{\rm rec}\le [r_v^*-r_v(t_a)]_+/\sigma_v.
 \label{eq:recovery_time}
\end{equation}
The positive rate must hold until that time.
\end{corollary}
\begin{IEEEproof}
The reflected integral begins at $R_v(t_a)$ and stays below $r_v^*$. If $z$ were to exceed $R_v$, the first crossing would occur below the cap. There $\dot R_v\ge b=\dot S\ge\dot z$ almost everywhere. The positive-part comparison for $[z-R_v]_+$ excludes the crossing and gives $R_v\ge z$. The position rate also satisfies $\dot r_p\ge G_p-A_p\ge0$ on its layer, so the position reserve remains positive. Continuation follows as in Theorem~\ref{thm:two}. Integrating the constant negative rate yields \eqref{eq:loss_time}; integrating a positive rate up to the cap yields \eqref{eq:recovery_time}.
\end{IEEEproof}
For repeated losses beginning with $R_v=r_v^*$, suppose each loss consumes at most $\bar Q<r_v^*$ and each recovery interval lasts at least $\bar Q/\sigma_v$. The reserve then returns to the cap before the next loss, so the argument applies by induction when the onset times tend to infinity. A smaller initial reserve requires a correspondingly smaller first loss budget. The recovery estimate concerns trajectories that remain inside both error bounds.

\subsection{Fixed Feedback and Initial-State Bounds}\label{subsec:fixed}
Let $\ell_v^0$ bound the directional rate for residual allowance $\delta_0$ and gain lower bound $b_0$. Keep the command map, nominal drift, and coordinates fixed, while allowing residual $\delta_0+\varepsilon_f(t,\bx)$ and lower gain $b_0-\varepsilon_B(t)>0$. Choose an integrable envelope
\begin{equation}
 c_\Delta(t)\ge\sup_{\D(t),\,0\le r_v\le r_v^*}
 \frac{2q_v}{\rho_2^2}
 [\varepsilon_f(t,\bx)+\eta_L(t)\varepsilon_B(t)a_v].
 \label{eq:uncertainty_cost}
\end{equation}
\begin{corollary}[Additional uncertainty under the same command]\label{cor:fixed}
Under the conditions of Theorem~\ref{thm:two} for the enlarged uncertainty class, suppose $G_p\ge A_p$ and
\begin{equation}
 \int_{t_a}^{t_b}[-\ell_v^0(t)]_+dt+
 \int_{t_a}^{t_b}c_\Delta(t)dt<R_v(t_a).
 \label{eq:fixed_test}
\end{equation}
Then the unchanged feedback preserves both bounds.
\end{corollary}
\begin{IEEEproof}
For the enlarged uncertainty class, \eqref{eq:rv_identity} gives $\dot r_v\ge\ell_v^0-c_\Delta$. The inequality $[-\ell_v^0+c_\Delta]_+\le[-\ell_v^0]_++c_\Delta$ then bounds the accumulated loss by the two integrals in \eqref{eq:fixed_test}. The position estimate is unchanged. Applying Theorem~\ref{thm:two} proves preservation of both reserves.
\end{IEEEproof}
The preceding condition uses bounds over the full candidate region. For a given onset state, the following estimate uses a smaller region over a finite interval.

\paragraph{Finite-horizon bound from the onset state}\label{subsec:onset_region}
Write $\bm\nu=\bv-\dot\bpd$ and $\kappa=k_p/D_p$. For an analysis constant $\lambda>0$, define
\begin{equation}
 \bm\zeta=\bm\nu+\lambda\bep,\quad
 W=\bigl(\|\bm\zeta\|^2+\lambda^2\|\bep\|^2\bigr)^{1/2}.
 \label{eq:onset_W}
\end{equation}
The analysis uses $W$ and $\lambda$ with the original feedback \eqref{eq:controller}. On the working region, suppose $\bB=b\bI_m+\bm E$, with $0<\underline b\le b\le\overline b$, $\|\bm E\|\le\varepsilon_E$, and $0\le\eta_-\le\eta(t)\le\eta_+$. Set $A_-=\eta_-\underline bU$ and $A_+=\eta_+\overline bU$. The bounds needed below are
\begin{equation}
 h\ge\frac{U(\rho_{2,+}^2+\epsilon_v)}{m_*k_v},\quad
 \|\bff-\ddot\bpd+\bd\|+\eta_+U\varepsilon_E\le M.
 \label{eq:onset_data}
\end{equation}
For $W\le W_*$, let $d_\kappa\ge|\kappa-\lambda|$. Choose $0<c<1$ such that
\begin{equation}
 \frac{d_\kappa^2}{\lambda^2}\le\frac{c^2}{1-c^2},\quad
 \frac{W_*}{\lambda}<\rho_{1,-},\quad
 \frac{W_*}{\sqrt{1-c^2}}<\rho_{2,-}.
 \label{eq:onset_domain}
\end{equation}

The conditions in \eqref{eq:onset_domain} ensure that $W\le W_*$ lies inside both error bounds. To bound the input contribution, we distinguish $\|\bm\zeta\|/W\le c$ and $\|\bm\zeta\|/W\ge c$. Put $a_\lambda=\lambda(2c^2-1)$ and define, for $w>0$,
\begin{equation}
\begin{aligned}
 F_-(w)&=a_\lambda w+(M+A_+)c,\\
 F_+(w)&=\min\{\lambda w+M,\\
 &\quad\max\{a_\lambda w+Mc,\lambda w+M-A_-\}
 +\frac{A_-h}{2w}\},\\
 F(w)&=\max\{F_-(w),F_+(w)\}.
\end{aligned}\label{eq:onset_F}
\end{equation}
\begin{proposition}[An initial-state duration bound]\label{prop:onset_duration}
Suppose Assumptions~\ref{ass:model}--\ref{ass:reference} and the above bounds hold on $[t_f,t_f+T]$. Assume the region $W\le W_*$ lies in a compact subset of $\mathcal X$, and $W(t_f)\le W_0$ with $0<W_0<W_b\le W_*$. If $F>0$ on $[W_0,W_b]$ and
\begin{equation}
 \int_{W_0}^{W_b}\frac{dw}{F(w)}>T,
 \label{eq:onset_time}
\end{equation}
then the solution continues through $t_f+T$ with $W(t)<W_b$ and
\begin{equation}
 r_p(t)>1-\left(\frac{W_b}{\lambda\rho_{1,-}}\right)^2,\quad
 r_v(t)>1-\frac{W_b^2}{(1-c^2)\rho_{2,-}^2}.
 \label{eq:onset_margins}
\end{equation}
\end{proposition}
\begin{IEEEproof}
Let $q=\|\bev\|$ and $\bm r=\bff-\ddot\bpd+\bd+\eta\bm E\bu$, for which $\|\bm r\|\le M$. Using $\bev=\bm\zeta+(\kappa-\lambda)\bep$, differentiation of \eqref{eq:onset_W} gives
\begin{equation}
 \tfrac12\tfrac{d}{dt}W^2=
 \lambda(\|\bm\zeta\|^2-\lambda^2\|\bep\|^2)
 +\bm\zeta\T\bm r-
 \eta b a_v\frac{\bm\zeta\T\bev}{q}.
 \label{eq:onset_identity}
\end{equation}
At $q=0$, the input term vanishes. For $q>0$, the amplitude satisfies $A_-\min\{q/h,1\}\le\eta b a_v\le A_+$. Set $x=\|\bm\zeta\|/W$ for $W>0$. We estimate the input projection separately on $x\le c$ and $x\ge c$.

\emph{Case $x\le c$:} The input projection may increase $W$. Its contribution is at most $A_+\|\bm\zeta\|$, since $\eta b a_v\le A_+$. The other terms in \eqref{eq:onset_identity} then give $\dot W\le F_-(W)$.

\emph{Case $x\ge c$:} From \eqref{eq:onset_domain},
\begin{equation}
 \|\bm\zeta\|^2-(\kappa-\lambda)^2\|\bep\|^2
 \ge W^2\frac{x^2-c^2}{1-c^2}.
 \label{eq:onset_geometry}
\end{equation}
Using $2\bm\zeta\T\bev=q^2+\|\bm\zeta\|^2-(\kappa-\lambda)^2\|\bep\|^2$, the input projection is nonnegative in this region. Supplementary Appendix~C bounds it from below by
$A_-[W\sqrt{(x^2-c^2)/(1-c^2)}-h/2]$ for this contribution.
Inserting this bound into \eqref{eq:onset_identity} gives an upper rate that is convex in $x\in[c,1]$. The maximum occurs at an endpoint. Omitting the inward term gives a second upper bound. Their minimum is $F_+(W)$, so $\dot W\le F(W)$ in both regions.

If $W$ were to reach $W_b$ by $t_f+T$, consider its last preceding crossing of $W_0$. The bound $\dot W\le F(W)$ requires at least $\int_{W_0}^{W_b}dw/F(w)$ time between these crossings. By \eqref{eq:onset_time}, this time exceeds $T$. Thus $W<W_b$ on the stated interval. Finally, $\|\bep\|\le W/\lambda$ and $\|\bev\|\le W/\sqrt{1-c^2}$ yield \eqref{eq:onset_margins}. The compact-region condition permits continuation of the solution.
\end{IEEEproof}
For the spacecraft model, $b=1/m_s$, $\bm E=0$, and $\eta_- =\eta_+=\eta_s$ give $A_-=A_+=\eta_sU/m_s$. Matrix deviations enter $M$. Time-varying effectiveness requires both interval bounds because $F_-$ uses the largest adverse projection. The resulting estimate gives a lower bound on the time before either prescribed error boundary is reached.

\section{Input Requirements and Scalar Example}\label{sec:input}
To compare the preceding sufficient bounds with the input required by the plant, fix a physical constraint set $\mathcal C(t)$, a horizon, an initial state $\bx_0$, and an uncertainty class. Following the set-invariance formulation in \cite{Blanchini1999,Aubin1991,Miani2015}, let $\mathcal K_q(t)$ contain states from which a single measurable feedback $\bm w$, locally Lipschitz in the state and satisfying $\|\bm w\|\le q$, maintains $\mathcal C$ for every admitted uncertainty. For constant effectiveness, $\bm w=\eta\bu$. Define
\begin{equation}
 q_{\min}(\bx_0)=\inf\{q:\bx_0\in\mathcal K_q(0)\},
 \qquad \eta_{\rm crit}=q_{\min}/U.
 \label{eq:qmin}
\end{equation}
The infimum refers to the stated initial state and uncertainty class over the chosen horizon. At the limiting input level, the physical boundary may be reached.

For the radial feedback, the directional rate gives
\begin{equation}
 \eta_B(\bm\vartheta)=
 \operatorname*{ess\,sup}_{t}\ 
 \sup_{\D(t),\,0\le r_v\le r_v^*}
 \frac{[N_v]_+}{b_-q_va_v},
 \label{eq:etaB}
\end{equation}
where $N_v$ is defined above \eqref{eq:authority_rate}, and $\bm\vartheta$ collects design parameters. Its denominator is positive on the velocity layer. When $\eta_L(t)\ge\eta_B(\bm\vartheta)$ on the fixed horizon and the position-layer rate is nonnegative, the feedback retains both bounds. Using the full amplitude $a_v$ gives a threshold at most equal to the threshold obtained with $a_v^0$. The thresholds coincide when $a_v^0=U$ throughout the layer.

Let $\Theta_B$ be a nonempty compact set of design parameters. Its members have positive initial reserves, nonnegative position-layer rates, $\eta_B\le\bar\eta$, and satisfy the required compact-domain condition. Assume its normalized layer data are continuous in the parameters. Then $\eta_B$ is lower semicontinuous as a supremum of continuous ratios, so $\eta_B^\star=\min_{\Theta_B}\eta_B$ is attained. If $\D(t)\subseteq\mathcal C(t)$,
\begin{equation}
 \eta_{\rm crit}\le\eta_B^\star.
 \label{eq:ordering}
\end{equation}
For fixed $\eta$, any initial state whose physical constraint is maintained under Theorem~\ref{thm:two} or Proposition~\ref{prop:onset_duration} also lies in $\mathcal K_{\eta U}$ for the same horizon and uncertainty class.

At a point on the auxiliary boundary, $L_v^{\max}<0$ means that an admissible disturbance gives an outward reserve rate for every input bounded by $U$. This is a local obstruction to maintaining the auxiliary constraint. The integral condition, by comparison, gives sufficient input levels for preserving both error bounds. Their difference reflects the auxiliary constraint, the uniform estimates, and the selected parameter family.

A scalar position-velocity system allows the plant requirement and the feedback condition to be calculated for the same physical limits.
\begin{proposition}[A calculation with bounded uncertainty]\label{prop:example}
Consider $\dot p=v$, $\dot v=B(t,p,v)w+a(t,p,v)$ with $|a|\le D_0$, $b_0\le B\le b_1$, and the closed physical box $|p|,|v|\le1$. Assume the local regularity above and $D_0,b_0,b_1>0$. For $|p_0|<1$, $0<v_0\le1$, and an infinite horizon,
\begin{equation}
 \eta_{\rm nec}=\frac{D_0}{b_0U},\qquad
 \eta_{\rm crit}=\frac{D_0+v_0^2/[2(1-p_0)]}{b_0U}.
 \label{eq:scalar_critical}
\end{equation}
Use $f_0=0$, $\delta=D_0$, $U=3$, $\rho_1=1$, $\rho_2=0.2$, and
\begin{equation}
\begin{gathered}
 \epsilon_p=1,\quad\epsilon_v=0.04,\quad m_*=1,\quad\beta=0.5,\\
 \nu_v=0.05,\quad\delta_v=0.01,\quad r_p^*=0.1,\quad r_v^*=0.4,\\
 (k_p,k_v)\in[0.27,0.35]\times[2,3]=\Theta_B .
\end{gathered}\label{eq:parameters}
\end{equation}
Then $G_p>A_p$ and
\begin{equation}
 \eta_B(k_p,k_v)=\frac{D_0+3k_p(0.2+k_p)}{3b_0}.
 \label{eq:example_eta}
\end{equation}
Its minimum occurs at $k_p=0.27$ for every stated $k_v$.
\end{proposition}
The stopping-distance argument and the calculation of the rate bound are given in Supplementary Appendix~A.

At the limiting input level the trajectory may touch a position face. Every larger input level admits a smaller invariant interval containing the initial state and braking path. Thus the same threshold is the infimum for strict preservation.

\TabRef{tab:thresholds} uses $(p_0,v_0)=(0,0.1)$. Its second column permits $|f|\le0.01$, $|d|\le0.05$, and $B\in[0.9,1.1]$. Both columns have $r_p(0)=1$, $r_v(0)=0.75$. The norm estimate gives $\eta_B\le\eta_B^{\rm n}=A_v/G_v$. Because the basic term already saturates throughout the layer, these values are unchanged if the compensator retains its nominal residual allowance. The ratios compare the sufficient feedback levels with the plant threshold for the same scalar system and initial state.
\begin{table}[!t]\centering\footnotesize
\caption{Actuator effectiveness levels in the scalar example}\label{tab:thresholds}
\begin{tabular*}{0.96\columnwidth}{@{\extracolsep{\fill}}lcc@{}}\toprule
Quantity&$D_0=0.05,\ b_0=1$&$D_0=0.06,\ b_0=0.9$\\\midrule
$\eta_{\rm nec}$&0.01667&0.02222\\
$\eta_{\rm crit}$&0.01833&0.02407\\
$\eta_B^\star$&0.1436&0.1632\\
$\eta_B^{\rm n}$ at $k_p=0.27$&0.1853&0.2107\\
$\eta_B^\star/\eta_{\rm crit}$&7.831&6.78\\\bottomrule
\end{tabular*}
\end{table}
For the uncertain case, $A_v=4.407$ gives $T<0.09076$~s from the full reserve cap. From the same state, Theorem~\ref{thm:outage}, with $M_*=0.06$ and $\rho_1=1$, gives $0.1T+0.03T^2<1$, or $T<4.342$~s. The position-only bound is about $47.8$ times the two-bound duration. The latter also accounts for the auxiliary velocity requirement and uses a uniform rate estimate.

\section{Numerical Results}\label{sec:simulation}
\subsection{Spacecraft Model and Nominal Evaluation}
To illustrate the results, consider the following spacecraft relative-motion model:
\begin{equation}
 \dot\bp=\bv,\qquad \dot\bv=\bm g(\bp,t)+\frac{\eta(t)}{m_s}\bu+\bd,
 \label{eq:spacecraft}
\end{equation}
where $m_s=120$~kg is the spacecraft mass, $U=35$~N, and
\begin{equation}
\begin{aligned}
 \bm g&=-\mu_E\left(\frac{\bm r_T+\bp}{\|\bm r_T+\bp\|^3}
                    -\frac{\bm r_T}{\|\bm r_T\|^3}\right),\\
 \bm r_T&=R_o[\cos(\omega_ot),\sin(\omega_ot),0]\T .
\end{aligned}\label{eq:gravity}
\end{equation}
The constants are $R_o=6.90\times10^6$~m, $\mu_E=3.99\times10^{14}$~m$^3$/s$^2$, and $\omega_o=1.10\times10^{-3}$~rad/s, where $\bm r_T$ is the target position. The nominal evaluation uses $\bfo=\bm g$, $\bar\Delta_f=0$, $b_-=1/m_s$, and $m_*=m_s$.  The reference and initial state are
\begin{equation}
\begin{aligned}
 \bpd(t)&=\begin{bmatrix}7\cos(0.105t)\\5\sin(0.105t)\\1.8\sin(0.210t)\end{bmatrix},\\
 \bp(0)&=[7.08,-0.04,0.03]\T,\\
 \bv(0)&=[-0.012,0.540,0.369]\T,
\end{aligned}\label{eq:reference_num}
\end{equation}
with positions in metres and velocities in metres per second. The radii are $\rho_1=0.32+0.78e^{-0.105t}$ and $\rho_2=0.26+0.56e^{-0.1t}$. Effectiveness takes values $1,0.80,0.58,0.30,0.78,0.44$, changing at $4.5,9,10,11,13.5$~s. The disturbance combines sinusoidal acceleration and smooth pulses near $10$ and $14.5$~s. Its expression and the nominal-envelope calculation are given in the supplementary numerical specification. The proposed controller uses $k_p=0.8056$, $k_v=5.819$, $\epsilon_p=0.035$, $\epsilon_v=0.025$, $\beta=0.64$, $\nu_v=0.04564$, and $\delta_v=0.02129$. Its other design parameters and the redesigned profiles are listed in Supplementary Appendix~B. The radial controller is denoted RS-PPC. The comparison uses interval-distance PPC (ID-PPC) \cite{Guo2022}, traditional PPC, and input-constrained funnel control (ICFC) \cite{Berger2024} under the same spacecraft model and $35$~N input limit.

\subsection{Margins, Loss Severity, and Redesigned Bounds}\label{subsec:simulation_representative}
\begin{figure}[!t]\centering
\includegraphics[width=.86\columnwidth]{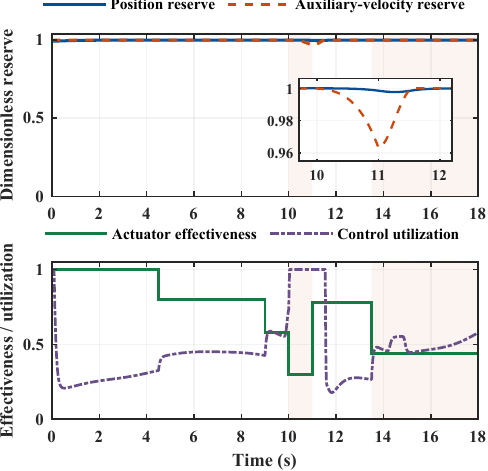}
\caption{Error reserves, effectiveness, and command utilization.}\label{fig:reserves}
\end{figure}

\begin{figure}[!t]\centering
\includegraphics[width=.86\columnwidth]{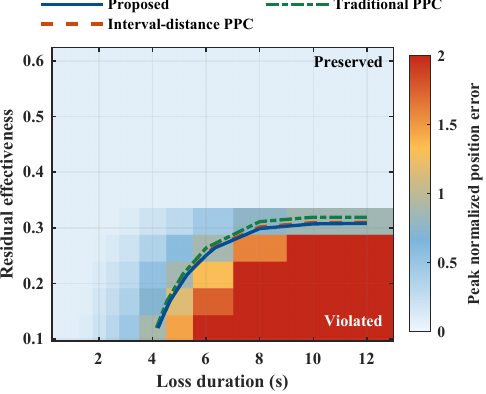}
\caption{Position-error peaks over the loss-duration grid.}\label{fig:severity}
\end{figure}

\FigRef{fig:reserves} shows the position and auxiliary velocity reserves and the ratio $\|\bu\|/U$. The minimum reserves are $0.9926$ and $0.9631$ over $0$--$18$~s, corresponding to $8.58\%$ and $19.21\%$ of the prescribed radii. The input respects $35$~N. \FigRef{fig:severity} shows the peak normalized position error across the loss-duration grid. The unit level of the normalized position error separates the passing and failing cases. On $I=[10,11]$~s, the prescribed layer widths $r_p^*=0.18$ and $r_v^*=0.30$ give $G_p\approx13.44$, $A_p\approx1.749$, $G_v\approx1.047$, and $A_v\approx3.075\times10^4$~s$^{-1}$ from \eqref{eq:uniform}. Thus $G_p>A_p$, while $\eta_B^{\rm n}\approx2.937\times10^4$. For effectiveness $0.30$, $R_v(10)=0.30$ and the uniform adverse integral is $Q_v\approx3.075\times10^4$. Equation~\eqref{eq:loss_time} gives $T<9.754\times10^{-6}$~s.  The region-wide estimate gives a duration much shorter than the one-second loss. Its large adverse term comes from bounding the virtual-velocity derivative over the full candidate region. Proposition~\ref{prop:onset_duration} uses the onset state to obtain a longer bound.

\begingroup
For the onset state recorded at $10$~s, $W(10)=0.005226$~m/s. With the unchanged command, $\eta_s=0.30$, $U=35$~N, and $\|\bd\|\le0.0895$~m/s$^2$, choose $\lambda=1.45$~s$^{-1}$, $W_0=0.006$~m/s, $W_*=0.280$~m/s, $W_b=0.276$~m/s, $c=0.601$, $h=0.01215$~m/s, and $M=0.1816$~m/s$^2$. An outward-rounded calculation of \eqref{eq:onset_time} gives a lower crossing time exceeding $1.008$~s. Proposition~\ref{prop:onset_duration} therefore yields
\begin{equation}
 r_p(t)>0.8867,\qquad r_v(t)>0.4015,
 \quad 10\le t\le11~\mathrm{s}.
 \label{eq:onset_result}
\end{equation}
The same command also admits $\|\bff-\bm g\|\le0.003$~m/s$^2$ and $\|\bB-\bI_3/120\|\le0.08/120$ on this interval. Including the gain deviation in the acceleration bound gives $r_p>0.8724$ and $r_v>0.3134$. Both cases stay above $r_p^*=0.18$ and $r_v^*=0.30$ for the initial set $W(10)\le0.006$~m/s. The enlarged case uses $c=0.6103$, $W_b=0.293$~m/s, and $W_*=0.294$~m/s, with the same $\lambda$, $W_0$, and $h$. These parameters enter the comparison estimate. The full parameter sets and rational lower sums are given in Supplementary Appendix~C.
\endgroup


\begin{figure}[!t]\centering
\includegraphics[width=.86\columnwidth]{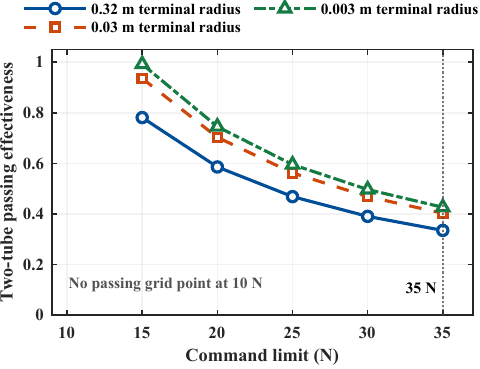}
\caption{Passing effectiveness after redesign for each profile.}\label{fig:budget}
\end{figure}

The three curves in \FigRef{fig:budget} correspond to feedbacks designed separately for their position bounds. The terminal radii are $0.32$, $0.03$, and $0.003$~m, with respective decay rates of $0.105$, $0.30$, and $0.40$~s$^{-1}$. Each test requires both the position and auxiliary velocity bounds to hold. \TabRef{tab:budget} lists the failed and passed effectiveness levels at $35$~N. The passing endpoints were checked with a smaller integration step.
\begin{table}[!t]\centering\footnotesize
\caption{Approximate failed and passed effectiveness at $35$ N}\label{tab:budget}
\begin{tabular*}{0.96\columnwidth}{@{\extracolsep{\fill}}lcc@{}}\toprule
Terminal radius (m)&Failed level&Passed level\\\midrule
0.32&0.3344&0.3352\\
0.03&0.4016&0.4031\\
0.003&0.425&0.4266\\\bottomrule
\end{tabular*}
\end{table}

With $\bfo=\bm0$, the bounds-only implementation was also tested against an unmodeled drift of $0.003$~m/s$^2$ and an $8\%$ input-gain variation. Its minimum position and auxiliary velocity reserves are $0.9926$ and $0.9664$, respectively, with $E_{\rm RMS}=0.00635$~m and $J_u=5569$~N$^2$s. The input remains within $35$~N. A run using the nominal model on the same uncertain plant gives $E_{\rm RMS}=0.00635$~m.

\subsection{Tracking and Input Effort}
\begin{figure}[!t]\centering
\includegraphics[width=.86\columnwidth]{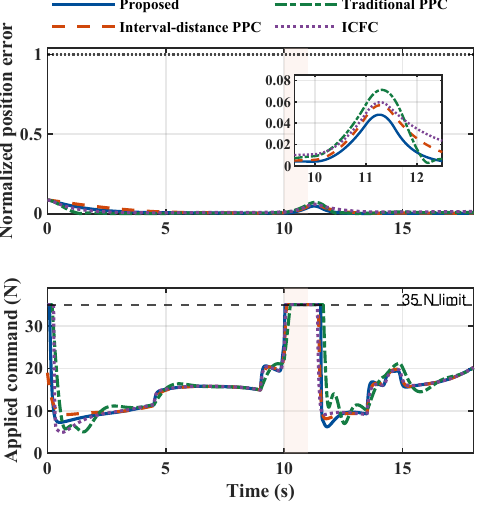}
\caption{Tracking errors and applied commands.}\label{fig:tracking}
\end{figure}

\FigRef{fig:tracking} compares $\chi_p=\|\bep\|/\rho_1$ and the actual command. Define
\begin{equation}
\begin{aligned}
 E_{\rm RMS}&=\left(\frac1{13.5}\int_{4.5}^{18}\|\bep(t)\|^2dt\right)^{1/2},\\
 J_u&=\int_0^{18}\|\bu(t)\|^2dt,
\end{aligned}\label{eq:metrics}
\end{equation}
where $E_{\rm RMS}$ is the RMS position error and $J_u$ is squared-command effort. \TabRef{tab:comparison} reports the four methods. RS-PPC gives the lowest RMS error ($0.00676$~m), while ID-PPC has the lowest squared-command effort ($5220$~N$^2$s). ICFC gives $E_{\rm RMS}=0.01064$~m and $J_u=5425$~N$^2$s. ICFC adjusts its funnel boundary using model-free feedback, whereas the nominal RS-PPC command uses $\bfo=\bm g$ and a known disturbance bound. All four inputs satisfy the $35$~N limit in this test. Supplementary Appendix~B lists the comparison settings.
\begin{table}[!t]
\centering
\footnotesize

\caption{RMS position error and squared-command effort}\label{tab:comparison}
\begin{tabular*}{0.96\columnwidth}{@{\extracolsep{\fill}}lcc@{}}
\toprule
Method & $E_{\rm RMS}$ (m) & $J_u$ (N$^2$s)\\
\midrule
RS-PPC & \textbf{0.00676} & 5416\\
ID-PPC & 0.00887 & \textbf{5220}\\
Traditional PPC & 0.01037 & 5801\\
ICFC & 0.01064 & 5425\\
\bottomrule
\end{tabular*}

\end{table}

\subsection{Complete Loss and Boundary Contact}\label{subsec:outage_num}
\begin{figure}[!t]\centering
\includegraphics[width=.93\columnwidth]{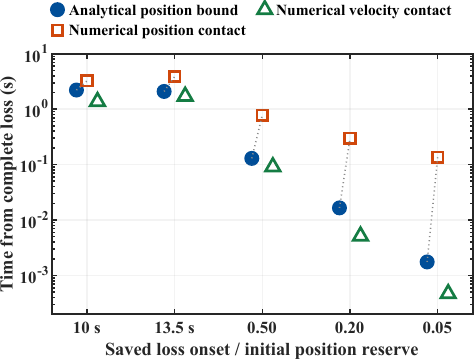}
\caption{Analytical position bound and numerical contact times for the position and auxiliary-velocity bounds.}\label{fig:complete}
\end{figure}

\FigRef{fig:complete} compares the conservative position-preservation duration with numerical contacts at the physical position and auxiliary velocity boundaries. From the nominal states at $10$ and $13.5$~s, the conservative physical-position times are approximately $2.214$ and $2.089$~s. The corresponding two-error-bound durations from Corollary~\ref{cor:recovery} are $7.704\times10^{-6}$ and $1.105\times10^{-5}$~s. In the uncontrolled trajectories, the auxiliary velocity boundary is first reached after $1.374$ and $1.711$~s, and the physical position boundary after $3.245$ and $3.878$~s, respectively. Additional onset states vary the physical position margin and error velocity. Their parameters appear in Supplementary Appendix~B.

\begin{figure}[!t]\centering
\includegraphics[width=.86\columnwidth]{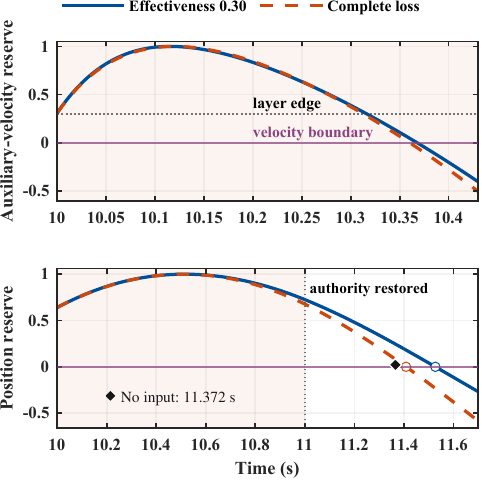}
\caption{Auxiliary and position margins in the boundary challenge. The diamond marks position contact under continued loss.}\label{fig:challenge}
\end{figure}

In \FigRef{fig:challenge}, both branches start at $t_f=10$~s with $r_p(t_f)=0.64$ and $r_v(t_f)=0.31$. The analysis layers are $r_p^*=0.18$ and $r_v^*=0.30$. For the complete-loss branch, the vector test of Theorem~\ref{thm:outage} gives $T<1.103$~s, and its continuous-time envelope remains inside the physical position bound over the tested $1$~s loss. The auxiliary boundary is reached after $0.3604$~s. The physical position boundary is reached $1.409$~s after loss onset, following actuation restoration at $11$~s.  Under continued complete loss, position contact occurs at $11.37$~s. With actuation restored, contact occurs at $11.41$~s, a delay of approximately $0.0375$~s. The partial-loss branch, with effectiveness $0.30$, reaches the auxiliary boundary after $0.3680$~s. Supplementary Appendix~B gives the common initial state and feedback parameters. The auxiliary reserve reaches zero before restoration. Corollary~\ref{cor:recovery} concerns recovery starting with both reserves positive.

\section{Conclusion}

We studied how prescribed tracking bounds can be maintained when actuator effectiveness decreases under a fixed input limit. For projected radial feedback, the applied input determines lower bounds on the position and auxiliary-velocity reserve rates. Integrating these rates gives conditions for retaining both bounds during partial loss and for restoring the auxiliary reserve. A bound based on the state at loss onset also admits bounded gain perturbations. Under complete loss, the position estimate gives a duration over which the physical constraint holds, even when the auxiliary bound has been reached. Finally, the scalar stopping-distance calculation compares the input required by the plant with a sufficient level for the stated feedback family. The spacecraft results illustrate the two kinds of bounds and their relation to observed boundary contacts.

\section*{Acknowledgment}
OpenAI's ChatGPT assisted with language editing and LaTeX preparation. The authors are responsible for the final arguments and results.

\end{document}